\documentclass[aps,pre,twocolumn,showpacs,showkeys,groupedaddress,amsmath,twoside]{revtex4-1}

\usepackage{amsmath,amsfonts,amssymb,subfigure}
\usepackage{graphicx,dsfont,natbib,verbatim,color,array}

  \newcommand{\esp   }[1]{\mathds{E}[#1]}
  
  \newcommand{\pr    }[1]{\mathds{P}[#1]}

\renewcommand{\Pr    }[1]{\mathds{P}\!\left[#1\right]}
  
\newcommand{\dd     }[1]{\mathrm{d}#1} 

\newcommand{\Xm}{X^{\mathrm{\scriptscriptstyle{(-)}}}}
\newcommand{\Xp}{X^{\mathrm{\scriptscriptstyle{(+)}}}}
\newcommand{\Xpm}{X^{\mathrm{\scriptscriptstyle{(\pm)}}}}
\newcommand{\vev}[1]{\langle #1 \rangle}

  \newcommand{\cop  }[2][]{C_{#1}#2}
  
  \newcommand{\copN }[2][]{\mathcal{C}_{#1}#2} %
  \newcommand{\CopN }[2][]{\mathcal{C}_{#1}\!\left(#2\right)} %

\newcommand{\ribracket}{] \mspace{-3 mu} ]}
\newcommand{\libracket}{[\mspace{-3 mu} [}

\begin{document}

\title{Copulas and time series with long-ranged dependences}

\author{R\'emy~Chicheportiche}
\email{remy.chicheportiche@graduates.centraliens.net}
\author{Anirban Chakraborti}
\email{anirban.chakraborti@ecp.fr}
\affiliation{
Chaire de finance quantitative, \'Ecole Centrale Paris, 92\,295 Ch\^atenay-Malabry, France
}

\date{\today}

\begin{abstract}
We review ideas on temporal dependences and recurrences in discrete time series 
from several areas of natural and social sciences. 
We revisit existing studies and redefine the relevant observables in the language of copulas (joint laws of the ranks).
We propose that copulas  provide an appropriate mathematical framework to study non-linear time dependences and related concepts --- like aftershocks, Omori law, recurrences, waiting times. We also 
critically argue using this global approach that previous phenomenological attempts involving only a long-ranged autocorrelation function lacked complexity in that they were essentially mono-scale.
\end{abstract}


\maketitle

\section{Introduction}

A thorough understanding of the occurences and statistics of extreme events is crucial in fields like seismicity, finance, astronomy, physiology, etc.\ \cite{Kotz2000,Embrechts1997}. The analyses of extreme events plays a pivotal role every time an addressed problem has a stochastic nature, since the rare extreme events can have rather strong or drastic consequences--- making it widely useful. One theoretical motivation for studying extreme events in a particular field like finance, is to account for the observed fat tails of log-returns (deviation from the Normal distribution in the tails) of stock prices \cite{chakraborti2011econophysics1}. A more practical motivation is that the extreme events such as ``market crashes'' or ``shocks'', pose a substantial risk for investors, even though these events are rare and do not provide enough data for reliable statistical analyses \cite{politi2012near}. It has been observed that common financial shocks are relatively smaller in magnitudes of volatility, the duration, and the number of stocks affected. However, the extremely large and infrequent financial crashes, such as the Black Monday crash, have significant ``aftershocks'' that can last for many months. This observation is very similar to the ``dynamic relaxation'' of the aftershock cascade following an earthquake. 
Hence, it is meaningful to also ask the general scientific question: How is the dynamics of a ``complex'' system, such as an earthquake fault \cite{bak2002unified,Bunde20031,PhysRevLett.92.108501,molchan2005interevent,altmann2005recurrence,saichev2007theory,sornette2008solution,blender2008extreme} or a financial market \cite{yamasaki2005scaling,wang2006scaling,PhysRevE.76.016109,Sazuka20092839,ren2010recurrence}, affected when the system undergoes an extreme event? The statistics of return intervals between extreme events is a powerful tool to characterize the temporal scaling properties of the observed time series and to estimate the risk for such hazardous events like earthquakes or financial crashes. Evaluating the return time statistics of extreme events in a stochastic process, is one of the classical problems in probability theory. 

Earlier, from an analysis of the probability density functions (PDF) of waiting times for earthquakes, Bak et al. \cite{bak2002unified} had suggested a unified scaling law combining the Gutenberg-Richter law, the Omori law, and the fractal distribution law in a single framework. This global approach was later extended by Corral \cite{corral2003local,corral2004universal}, who proposed the existence of a universal scaling law for the PDF recurrence times between earthquakes in a given region. This is useful because, due to the scaling properties, it is possible to analyse the statistics of return intervals for different thresholds by studying only the behavior of small fluctuations occurring very frequently, which have much better statistics and reliability than those of the rare extreme large flucutations. It also reveals a spatiotemporal organization of the seismicity, as suggested by \citet{saichev2007theory}.

In this paper, we review the ideas on temporal dependences and recurrences in discrete time series 
from several areas of  earthquakes, etc.\~(natural sciences) and  financial markets (social sciences). 
We revisit the existing studies, cited above, and redefine the relevant observables in the mathematical language of ``copulas''. We propose that copulas is a very general and appropriate framework to study non-linear time dependences and related concepts --- like aftershocks, Omori law, recurrences, waiting times. Our overall aim is to study several properties of recurrence times and the statistic of other observables (waiting times, cluster sizes, records, aftershocks) described in terms of the diagonal copula. We hope that these studies can shed light on the $n$-points properties of the process. We also critically argue that 
that previous phenomenological attempts involving only a long-ranged autocorrelation function,
lacked complexity in that they were essentially mono-scale.

\subsection*{The copula}
As a tool to study the --- possibly highly non-linear --- correlations between random variables, 
``copulas'', i.e.\ joint distributions of the ranks (see formal definition below), have long been used in actuarial sciences and finance to describe and model cross-dependences of assets, 
often in a risk management perspective \cite{embrechts2003modelling_art,embrechts2002correlation_art,malevergne2006extreme}.
Although the widespread use of simple analytical copulas to model multivariate dependences is more and more criticized \cite{mikosch2006copulas,chicheportiche2012joint}, 
copulas remain useful as a tool to investigate empirical properties of multivariate data \cite{chicheportiche2012joint}.

More recently, copulas have also been studied in the context of serial dependences in univariate time series, 
where they find yet another application range: just as Pearson's $\rho$ coefficient is commonly used to 
measure both linear cross-dependences and temporal correlations, 
copulas are well-designed to assess non-linear dependences both transversally or serially
\cite{beare2010copulas,ibragimov2008copulas,patton2009copula_art} --- we will speak of ``self-copulas'' in the latter case. 

\subsection*{Notations}
We consider a time series $\{X_t\}_{t=1\ldots T}$ of length $T$, 
 as a realization of a discrete stochastic process.
The joint cumulative distribution function (CDF) of $n$ occurrences ($1\leq t_1<\ldots<t_n<T$) of the process is 
\begin{equation}\label{eq:def_multiF}
    \mathcal{F}_{t_1,\ldots,t_n}(\mathbf{x})=\pr{X_{t_1}<x_{t_1},\ldots,X_{t_n}<x_{t_n}}.
\end{equation}
We assume that the process is stationary with a distribution $F$,
and a translational-invariant joint distribution $\mathcal{F}$ with long-ranged dependences, 
as is typically the case e.g.\ for seismic and financial data.

A realization of $X_t$ at date $t$ will be called an ``event'' when its value exceeds a threshold $\Xpm$:
``negative event'' when $X_t<\Xm$, and ``positive event'' when $X_t>\Xp$.
The probability $p_-$ of such a `negative event' is $F(\Xm)$, and similarly, 
the probability that $X_t$ is above a threshold $\Xp$ is the tail probability $p_+=1-F(\Xp)$.

\begin{figure}
\center
    \subfigure[$F(\Xp)=1-F(\Xm)=q$]{\label{fig:schema_b}\includegraphics[scale=0.55,trim=0   0 220 0,clip]{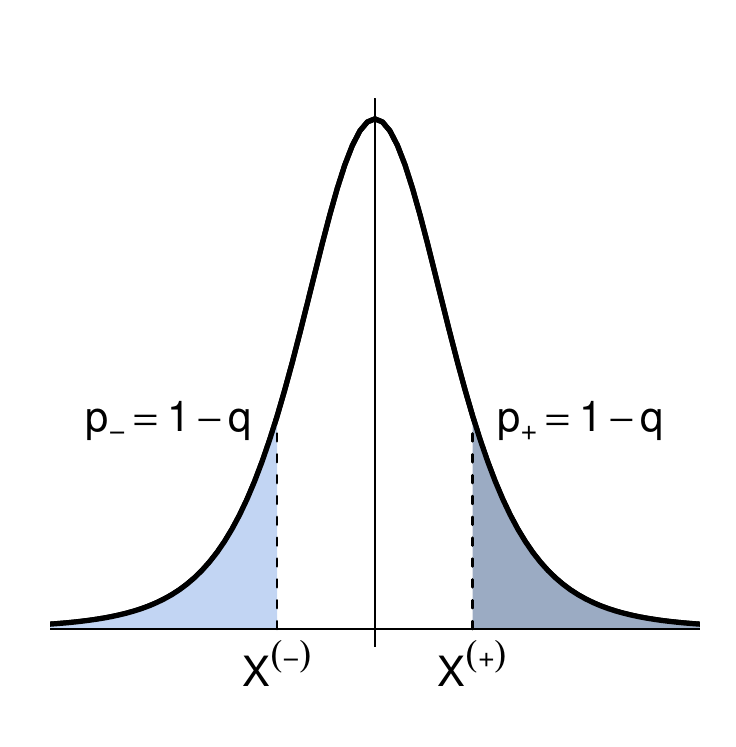}}\hfill
    \subfigure[$F(\Xp)=  F(\Xm)=q$]{\label{fig:schema_a}\includegraphics[scale=0.55,trim=220 0   0 0,clip]{images/schema.pdf}}
    \caption{Two possible definitions of events: 
            either $p_-$ and $p_+$ are probabilities of extremes (negative and positive, respectively),
            or only $p_+$ is a probability of extreme and $p_-=1-p_+$.}
	\label{fig:schema}
\end{figure}

If a unique threshold $\Xp=\Xm$ is chosen, then obviously $p_+=1-p_-$.
This is appropriate when the distribution is one-sided, typically for positive only signals,
and one wishes to distinguish between two regimes: extreme events (above the unique threshold), and regular events (below the threshold).
This case is illustrated schematically in Fig.~\ref{fig:schema_a}.
When the distribution is two-sided, it is more convenient to define, $\Xp$ as the $q$-th quantile of $F$, 
and $\Xm$ as the $(1\!-\!q)$-th quantile, for a given $q\in[\tfrac{1}{2},1]$,
so that $p_+=p_-=1-q$. 
This allows to investigate persistence and reversion effects in \emph{signed} extreme events,
while excluding a neutral zone of regular events between $\Xm$ and $\Xp$, see Fig.~\ref{fig:schema_b}

When the threshold for the recurrence is defined in terms of quantiles like above (a \emph{relative} threshold), 
stationarity is not needed theoretically but much wanted empirically as already said,
otherwise the height of the threshold might change every time. 
In contrast, when the threshold is set as a number (an \emph{absolute} threshold),
 there's no issue on the empirical side,
but the theoretical discussion makes sense only under stationary marginal.

The next section recalls several two-points and many-points properties of stationary processes,
and discusses associated measures of dependence in light of the copula.
This rather theoretical content is followed in Section~III by applications to financial data.
The definition and some properties of copulas are recalled in appendix, 
and the Gaussian case with long-ranged correlations is treated.

\section{Dependences in discrete-time processes}
We consider the case where the discrete times $t_n$ 
in the definition \eqref{eq:def_multiF} are equidistant (``regularly sampled'').

\subsection{Two-points dependence measures}
Typical measures of dependences in stationary processes are two-points expectations
that only involve one parameter: the lag $\ell$ separating the points in time.
For example, the usefulness of the linear correlation function
\begin{align}\label{eq:def_rho_2points}
    \rho(\ell)&=\esp{X_tX_{t+\ell}}-\esp{X_t}\,\esp{X_{t+\ell}}
\end{align}
is rooted in the analysis of Gaussian processes, as
those are completely characterized by their covariances, and
multi-linear correlations are reducible to all combinations of $2$-points expectations, according to Isserli's theorem.
Some non-linear dependences, like the tail-dependence for example \cite{embrechts2002correlation_art,malevergne2006extreme}, 
are however not expressed in terms of simple correlations, but involve the whole bivariate copula:
\begin{equation}\label{eq:biv_cop}
    \cop[\ell](u,v)=\mathcal{F}_{t,t+\ell}(F^{-1}(u),F^{-1}(v)),
\end{equation}
where $(u,v)\in[0,1]^2$.
${C}_\ell$ can be understood as the distribution of the marginal ranks $U=F(X_t), V=F(X_{t+\ell})$, and
contains the full information on bivariate dependence that is invariant under increasing transformations of the marginals.
For example, the conditional probability
\begin{equation}\label{eq:def_p_plusplus}
    p_{++}^{(\ell)}=\pr{X_{t+\ell}>\Xp|X_t>\Xp},
\end{equation}
which is a measure of \emph{persistence} of the ``positive'' events, 
can be written in terms of copulas, 
together with all three other cases of conditioning
\begin{subequations}\label{eq:allp}
\begin{align}
     p_{++}^{(\ell)}&=[{2p_+-1+\cop[\ell](1\!-\!p_+,1\!-\!p_+)}]/{p_+},\\
     p_{--}^{(\ell)}&= {\cop[\ell](p_-,p_-)}/{p_-},\\
     p_{-+}^{(\ell)}&=[{p_--\cop[\ell](p_-,1\!-\!p_+)}]/{p_-},\\
     p_{+-}^{(\ell)}&=[{p_--\cop[\ell](1\!-\!p_+,p_-)}]/{p_+}
\end{align}
\end{subequations}
where $p_{\pm\pm}^{(\ell)}$ and $p_{\mp\mp}^{(\ell)}$ are defined similarly to Eq.~\eqref{eq:def_p_plusplus} with obvious inequality sign choices.
When $\Xp=\Xm=0$ and $\ell=1$, this is exactly the definition of \citet{boguna2004conditional}, 
with accordingly $p_-=p_+=F(0)$, see Fig.~\ref{fig:schema}.
Note also that $p_{\pm\pm}^{(\ell)}$ and $p_{\pm\mp}^{(\ell)}$ are straightforwardly related to 
the so-called `tail dependence coefficients' \cite{chicheportiche2013phd}.

\begin{figure}
\center
    \subfigure[Gaussian copula]         {\label{fig:condprobGauss} \includegraphics[scale=.41]{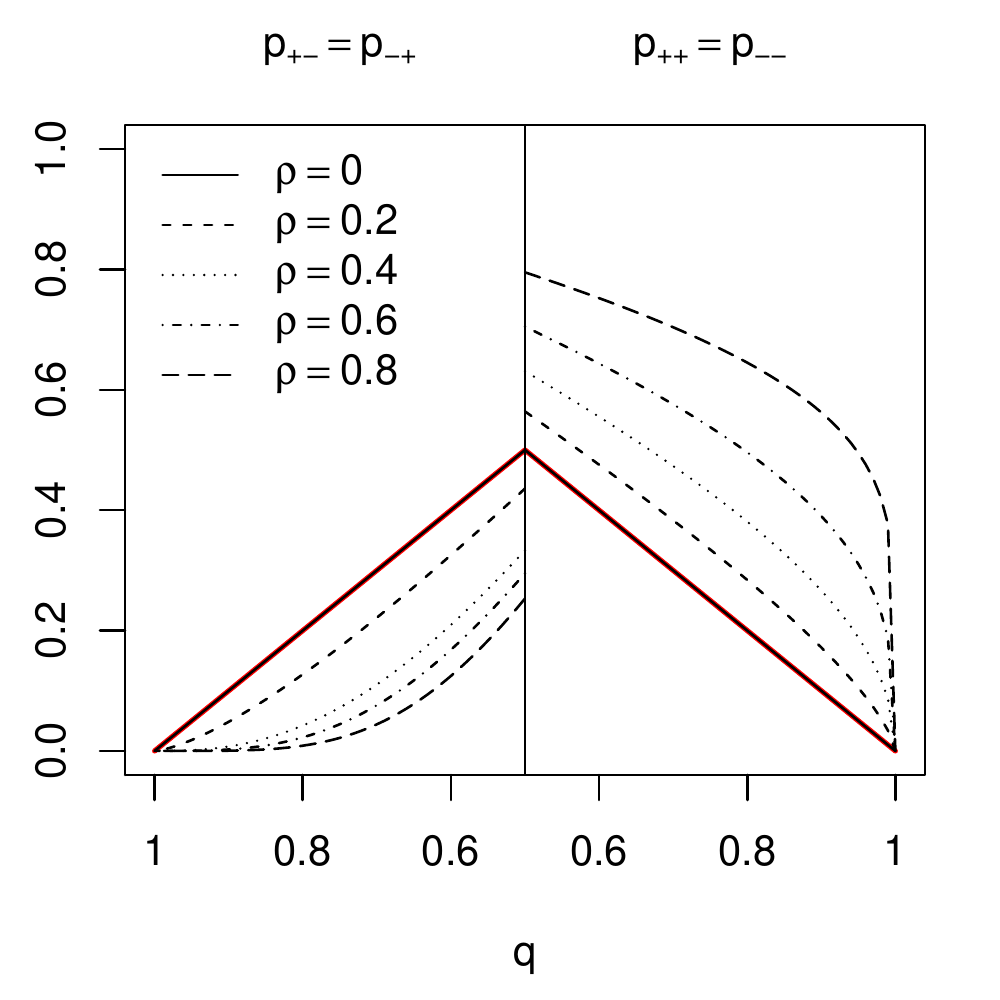}}\hfill
    \subfigure[Student copula ($\nu=5$)]{\label{fig:condprobStud5} \includegraphics[scale=.41]{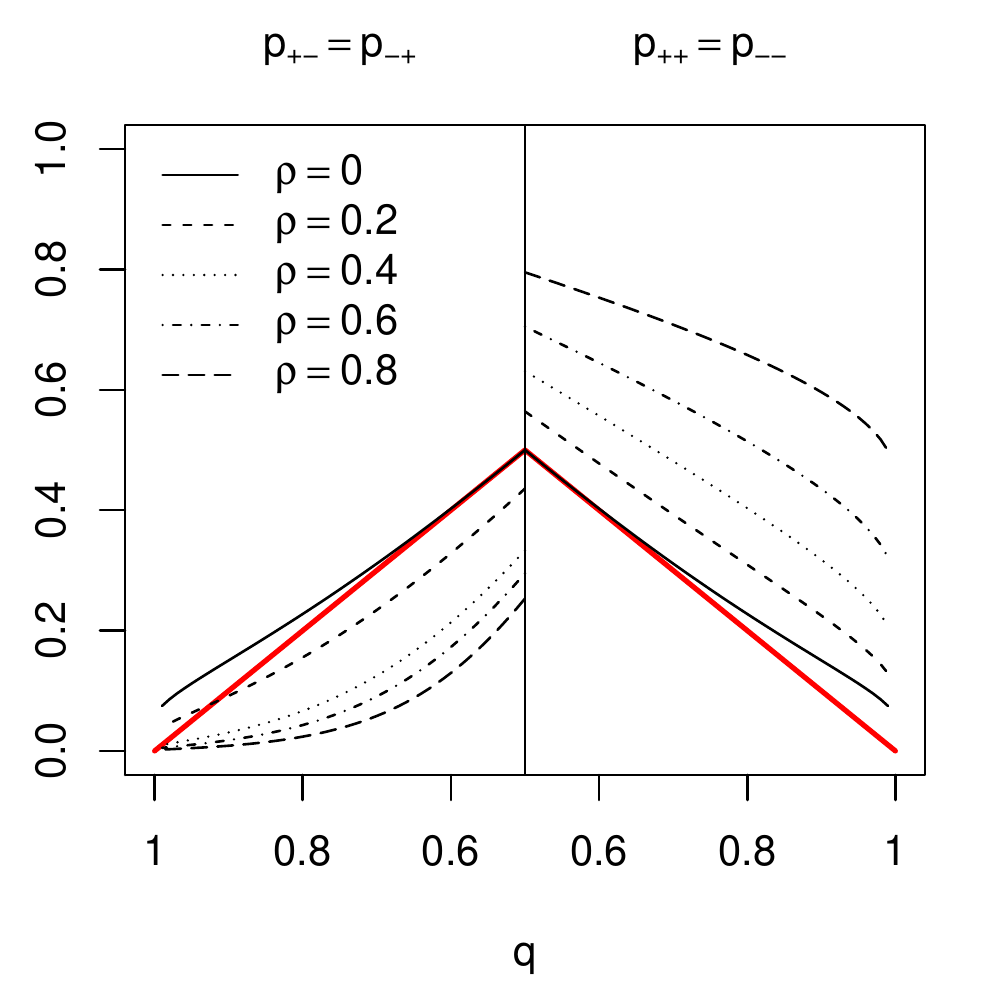}}
    \caption{Conditional probabilities $p_{\pm\mp}^{(\ell)}$ and $p_{\pm\pm}^{(\ell)}$ for different values of $\rho(\ell)$ with thresholds at $p_+=p_-=1-q$.
             The independent copula is shown in bold red.
             }\label{fig:condprob}
\end{figure}

As an example, consider the Gaussian bivariate copula of the pair $(X_t,X_{t+\ell})$,
whose whole $\ell$-dependence is in the linear correlation coefficient $\rho(\ell)$.
Fig.~\ref{fig:condprobGauss} illustrates the conditional probabilities \eqref{eq:allp} as a function of the threshold, when $p_+=p_-=1-q$.   
A similar plot for the Student copula (with $\nu=5$ degrees of freedom) is shown in Fig.~\ref{fig:condprobStud5}: 
the fatter tails of the joint distribution are responsible for the abnormal behavior of the conditional probabilities in the region $q=1$.
When $q=0.5$, the coefficients \eqref{eq:allp} are all equal to 
$$\frac{1}{2}+\frac{1}{\pi}\,\arcsin \rho(\ell)$$
for any elliptical copula \cite{chicheportiche2013phd}.

\subsubsection*{Aftershocks}

Omori's law characterizes the $\ell$ dependence of $p_{++}^{(\ell)}$,
i.e.\ the average frequency of events occurring $\ell$ time steps after a main event.
It was first stated in the context of earthquakes occurrences \cite{omori1895}, where this time dependence is power-law:
\begin{equation}\label{eq:omori}
    p_{++}^{(\ell)}= \lambda \cdot \ell^{-\alpha}.
\end{equation}
Notice that any dependence on the threshold must be hidden in $\lambda$ according to this description.
The average \emph{cumulated} number $N_\ell$ of these aftershocks until $\ell$ is thus 
\begin{equation}
    \vev{N_\ell}_+=\lambda \cdot \frac{\ell^{1-\alpha}}{1\!-\!\alpha},
\end{equation}
with in fact $\lambda\equiv p_+$ since, when $\alpha\to 0$, $N_{\ell}$ has no time-dependence, i.e.\ it counts independent events (white noise),
and $p_{++}^{(\ell)}$ must thus tend to the unconditional probability.

In order to give a phenomenological grounding to this empirical law also later observed in finance \cite{PhysRevE.76.016109,PhysRevE.82.036114},
\citet{lillo2003power} model the aftershock volatilities in financial markets as a decaying scale $\sigma(\ell)$ times an independent stochastic amplitude $r_\ell$ with CDF $\phi$.
As a consequence, $p_{++}^{(\ell)}\sim 1-\phi(\Xp/\sigma(\ell))$ and the power-law behavior of Omori's law results from 
(i) power-law marginal $\phi(r)\sim r^{-\gamma}$, {and}
(ii) scale decaying as power-law $\sigma(t)\sim t^{-\beta}$, 
so that relation \eqref{eq:omori} is recovered with $\alpha=\beta\gamma$.
The non-stationarity described by $\sigma$ is only introduced in a \emph{conditional} sense,
and might be appropriate for aging systems or financial markets, but we believe that
Omori's law  can be accounted for in a stationary setting and without necessarily having power-law distributed amplitudes.

{The scaling of $p_{++}^{(\ell)}$ with the magnitude of the main shock is encoded in the prefactor $\lambda\equiv p_+$,
which, for example, accounts for the exponentially distributed magnitudes of earthquakes (Gutenberg-Richter law \cite{Gutenberg21021936}).}
The linear dependence of $p_{++}^{(\ell)}$ on $p_+$ shall be reflected in the diagonal of the underlying copula:
\begin{equation}
    \cop[\ell](p,p)=p^2 \,\ell^{-\alpha},
\end{equation}
a prediction that can be tested empirically.

Note that Omori's law is a measure involving only the two-points probability.
In the next subsection, we show what additional information many-points probability can reflect.

\subsection{Multi-points dependence measures}
Although the $n$-points expectations of Gaussian processes reduce to all combinations of $2$-points expectations \eqref{eq:def_rho_2points},
{their full dependence structure is not reducible to the bivariate distribution}, 
unless the process is also Markovian (i.e.\ only in the particular case of exponential correlation).
Furthermore, when the process is not Gaussian, even the multi-linear correlations are irreducible.
In the general case, the whole multivariate CDF is needed, but many measures of dependence that we introduce below only 
involve the diagonal $n$-points copula:%
\footnote{We use a calligraphic $\mathcal{C}$ in order to make it clearly distinct from the bivariate copula discussed in the previous section.}
\begin{equation}\label{eq:def_cop_n}
    \CopN[n]{p}=\mathcal{F}_{t+\libracket 1,n\ribracket}(F^{-1}(p),\ldots,F^{-1}(p)),
\end{equation}
which measures the joint probability that all $n\geq 1$ consecutive variables $X_{t+1},\ldots,X_{t+n}$ are below 
the upper \mbox{$p$-th} quantile of the stationary distribution 
($p\in[0,1]$, and $t+\libracket 1,n\ribracket$ is a shorthand for $\{t\!+\!1,\ldots,t\!+\!n\}$).
Clearly, $\CopN[1]{p}=p$ and we set by convention $\CopN[0]{p}\equiv 1$.


Empirically, the $n$-points probabilities are very hard to measure due to the large noise associated
with such rare joint occurrences.
However, there exist observables that embed many-points properties and are more easily
measured, such as the length of sequences (clusters) of thresholded events,
and the recurrence times of such events, that we study next.

\subsubsection*{Recurrence intervals}
The probability $\pi(\tau)$ of observing a recurrence interval $\tau$ between two events 
is the conditional probability of observing a sequence of $\tau-1$ ``non-events'' bordered by two events:
\newcommand{\Rp}{\mathcal{X}^{\mathrm{\scriptscriptstyle{(+)}}}}
\begin{align}
    \pi(\tau)&=\pr{\Rp_{0;\tau}|X_{0}>\Xp},
\end{align}
where
\begin{align}
    \Rp_{t;\tau}&\equiv \big\{ X_{t+\libracket 1,\tau\libracket}<\Xp, X_{t+\tau}>\Xp \big\}
\end{align}
designates a sequence of `non-events' starting in $t$ and terminated by a `positive event' at $t+\tau$.
(We focus on positive events, but the recurrence of negative events can be studied with the substitution $X\to -X$,
and the case of recurrence in amplitudes with the substitution $X\to |X|$).
After a simple algebraic transformation flipping all `$>$' signs to `$<$', 
it can be written in the language of copulas as:
\begin{align}\label{eq:distrecint} 
        \pi(\tau)&=\frac{\CopN[\tau-1]{1\!-\! p_+}-2\,\CopN[\tau]{1\!-\! p_+}+\CopN[\tau+1]{1\!-\! p_+}}{p_+}.
\end{align}
The cumulative distribution
\begin{equation}\label{eq:cumul_Pi}
    \Pi(\tau)=\sum_{n=1}^\tau \pi(n)=1-\frac{\CopN[\tau]{1\!-\! p_+}-\CopN[\tau+1]{1\!-\! p_+}}{p_+}
\end{equation}
is more appropriate for empirical purposes, being less sensitive to noise.
These exact expressions make clear --- almost straight from the definition --- that 
  (i)~the distribution of recurrence times \emph{depends only on the copula} of the underlying process 
      and not on the stationary law, in particular its domain or its tails 
      (this is because we take a relative definition of the threshold as a quantile); 
 (ii)~\emph{non-linear} dependences are highly relevant in the statistics of recurrences, so that 
linear correlations can in the general case by no means explain alone the properties of $\pi(\tau)$; and 
(iii)~recurrence intervals have a \emph{long memory} revealed by the $(\tau\!+\!1)$-points copula being involved, 
so that only when the underlying process $X_t$ is Markovian will the recurrences themselves be memoryless.%
\footnote{It may be mentioned that in a non-stationary context, renewal processes are also able to produce independent consecutive recurrences \cite{PhysRevE.78.051113,Sazuka20092839}.}
 Hence, when the copula is known (Eq.~\eqref{eq:GaussCop} in appendix for Gaussian processes), 
 the distribution of recurrence times is characterized by the exact expression in Eq.~\eqref{eq:distrecint}.

The average recurrence time is found straightforwardly by summing the series
\begin{equation}\label{eq:avrecint}
{
    \mu_\pi=\vev{\tau}=\sum_{\tau=1}^{\infty}\tau\,\pi(\tau)=\frac{1}{p_+},
}
\end{equation}
and is \emph{universal} whatever the dependence structure.
This result was first stated and proven by Kac in a similar fashion~\cite{kac1947notion}. 
It is intuitive as, for a given threshold, the whole time series is the succession of a fixed number $p_+T$ of recurrences 
whose lengths $\tau_i$ necessarily add up to the total size $T$, so that $\vev{\tau}=\sum_i\tau_i/(p_+T)=1/p_+$.
Note that Eq.~\eqref{eq:avrecint} assumes an infinite range for the possible lags $\tau$, which is achieved either 
by having an infinitely long time series, or more practically when the translational-invariant copula is periodic at the boundaries of the time series, as
is typically the case for artificial data which are simulated using numerical Fourier Transform methods.
Introducing the copula allows to emphasize the validity of the statement even in the presence of non-linear long-term dependences,
as Eq.~\eqref{eq:avrecint} means that the average recurrence interval is \emph{copula}-independent.


More generally, the $m$-th moment can be computed as well by summing $\tau^m \pi(\tau)$ over $\tau$:
\[
    \vev{\tau^m}=\frac{1+\sum\limits_{\tau=1}^\infty\left[|\tau\!+\!1|^m-2\tau^m+|\tau\!-\!1|^m\right]\copN[\tau](1\!-\!p_+)}{p_+}.
\]
In particular, the variance of the distribution is
\begin{equation}\label{eq:secmon}
    \sigma_\pi^2\equiv\vev{\tau^2}-\mu_\pi^2=\frac{2}{p_+}\sum_{\tau=1}^{\infty}\CopN[\tau]{1\!-\!p_+}-\frac{1\!-\!p_+}{p_+^2},
\end{equation}
It is not universal, in contrast with the mean, 
and can be related to the average unconditional waiting time, see below.
Notice that in the independent case the variance $\sigma_\pi^2=(1-p_+)/p_+^2$ is not equal to the mean $\mu_\pi=1/p_+$,
 as would be the case for a continuous-time Poisson process, because of discreteness effects.

 It is important to notice that the main ingredient in the distribution of recurrence times \eqref{eq:cumul_Pi} is the copula 
 (i.e.\ the serial dependence structure) rather than the stationary distribution $F$, a finding already noted by~\citet{olla2007return},
 but which the current description highlights.
 The sensitivity to the extreme statistics of the process is in fact hidden in $p_+$, 
 but what matters more is the (possibly multi-scale) dependence structure $\mathcal{\cop}_{\tau}$. 

\subsubsection*{Conditional recurrence intervals, clustering}
The dynamics of recurrence times is as important as their statistical properties,
and in fact impacts the empirical determination of the latter.%
\footnote{Distribution testing for $\pi(\tau)$ involving Goodness-of-fit tests \cite{ren2010recurrence} 
should be discarded because those are not designed for dependent samples and rejection of the null
 cannot be relied upon. 
 See \citet{chicheportiche2011goodness} for an extension of GoF tests when some dependence is present. 
}
It is now clear, both from empirical evidences and analytically from the discussion on Eq.~\eqref{eq:distrecint}, 
that recurrence intervals have a long memory.
In dynamic terms, this means that their occurrences show some clustering. The natural question is then:
``Conditionally on an observed recurrence time, what is the probability distribution of the next one?''
This probability  of observing an interval $\tau'$ immediately following an observed recurrence time $\tau$
is
\begin{equation}
    \pr{\Rp_{\tau;\tau'}|\Rp_{0;\tau},X_0>\Xp}.
\end{equation}
Again, flipping the `$>$' to '$<$' allows to decompose it as
\[
     \frac{\copN[\tau-1;\tau'-1]{}-\copN[\tau;\tau'-1]{}-\copN[\tau-1;\tau']{}+\copN[\tau;\tau']{}}{\copN[\tau-1]{}-2\copN[\tau]{}+\copN[\tau+1]{}}
    -\frac{\pi(\tau+\tau')}{\pi(\tau)},
\]
where the $(\tau\!+\!\tau')$-points probability
\[
    \CopN[\tau;\tau']{p}=\mathcal{F}_{\libracket 0;\tau\!+\!\tau'\ribracket\backslash\{\tau\}}(F^{-1}(p),\ldots,F^{-1}(p))
\]
shows up.
Of course, this exact expression has no practical use, 
again because there is no hope of empirically measuring any many-points 
probabilities of extreme events with a meaningful signal-to-noise ratio.
We rather want to stress that non-linear correlations and multi-points dependences are relevant, 
and that a characterization of clustering based on the autocorrelation 
of recurrence intervals is an oversimplified view of reality.

\subsubsection*{Waiting times}

The conditional mean residual time to next event, when sitting $\tau$ time steps after a (positive) event, is
\begin{equation}\label{eq:residual_time}
    \vev{w|\tau}=\sum_{w=1}^\infty w\, \pi(\tau\!+\!w)=\frac{1}{p_+}\copN[\tau](1\!-\!p_+).
\end{equation}

One is often more concerned with unconditional waiting times,
which is equivalent to asking what the size $w$ of a sequence of `non-events' starting now will be,
\emph{regardless of what happened previously}. 
The distribution \mbox{$\phi(w)=\pr{\Rp_{0;w+1}}$} of these waiting times is equal to
\begin{align}
    \phi(w)&={\copN[w](1\!-\!p_+)-\copN[w+1](1\!-\!p_+)},
\end{align}
and its expected value is 
\begin{equation}
    \mu_{\phi}=\vev{w}=\sum_{w=1}^{\infty}\copN[w](1\!-\!p_+),
\end{equation}
consistently to what would be obtained by averaging $\vev{w|\tau}$  over
all possible elapsed times in Eq.~\eqref{eq:residual_time}.
%
From Eq.~\eqref{eq:secmon}, we have the following relationship between the variance of the distribution $\pi$
of recurrence intervals, and the mean waiting time:
\begin{equation}
    \sigma_\pi^2=\mu_\pi\,\big[2\mu_{\phi} +1 \big]-\mu_\pi^2
\end{equation}

\subsubsection*{Sequences lengths}
The serial dependence in the process is also revealed by the distribution of sequences sizes.
The probability that a sequence of consecutive negative events%
\footnote{We consider the case of ``negative'' events, i.e.\ those with $X_t<\Xm$
because it expresses simply in terms of diagonal copulas. 
The mirror case with ``positive'' events has the exact same expression but $\copN[n]$ must be inverted around the median.
For a symmetric $F$, this distinction is irrelevant.}, starting just after a `non-event',
will have a size $n$ is
\begin{equation}
    \psi(n)=\frac{\copN[n](p_-)-2\,\copN[n+1](p_-)+\copN[n+2](p_-)}{p_-\,(1-p_-)}
\end{equation}
and the average length of a random sequence 
\begin{equation}
{
    \mu_\psi=\vev{n}=\sum_{n=1}^{\infty}n\,\psi(n)=\frac{1}{1-p_-}
}
\end{equation}
is universal, just like the mean recurrence time.
This property rules out the analysis of \citet{boguna2004conditional} who claim to be able to distinguish 
the dependence in processes according to the average sequence size.

\subsubsection*{Record statistic}
We conclude this theoretical section on multi-points non-linear dependences
by mentioning that the diagonal $n$-points copula
$\copN[n]$ can be alternatively understood as the distribution of the maximum of 
$n$ realizations of $X$ {in a row}, since 
\[
    \Pr{\max_{\tau\leq n}\{X_{\tau}\}<F^{-1}(p)}=\Pr{X_{\libracket 1,n\ribracket}<F^{-1}(p)}
\]
is equal to $\copN[n](p)$. Thus,
studying the statistics of such ``local'' maxima in short sequences (see e.g.\ \cite{PhysRevE.73.016130}) 
can provide information on the multi-points properties of the underlying process.
The CDF of the {running} maximum, or \emph{record}, is $\copN[t](F(x))$ 
and the probability that $t>1$ will be a record-breaking time is the joint probability
\[
	R(t)=\Pr{\max_{\tau< t}\{X_{\tau}\}<X_t},
\]
which is \emph{irrespective of the marginal law} !


\newcolumntype{m}{>{$\displaystyle }c<{$}}

\begin{table}
\center
\begin{tabular}{||m|m||m|m||m|m||m||}
\hline
  \pi_+(\tau)             &\vev{\tau}_+     &\phi_+(w)       &\vev{w}_+        &\psi_-(n)              &\vev{n}_-    &R(t)\\\hline\hline
  (1\!-\!q)\,q^{\tau-1}   &\frac{1}{1\!-\!q}&(1\!-\!q)\,q^w  &\frac{q}{1\!-\!q}& q\,(1\!-\!q)^{n-1}    & \frac{1}{q} &\frac{1}{t}
\end{tabular}
\caption{Different probabilities introduced, with thresholds defined as $F(\Xp)=q=1-F(\Xm)$,
for the White Noise process.}
\end{table}

\section{Financial self-copulas}
\begin{table}
    \center
    \begin{tabular}{|lr||c|c|c|}
        Stock Index &Country        &From   &To     &$T$\\\hline\hline
        S\&P-500    &USA            &Jan.\ 02, 1970 &Dec.\ 23, 2011& 10\,615\\
        KOSPI-200   &S.    Korea    &Jan.\ 03, 1990 &Dec.\ 26, 2011&  5\,843\\
        CAC-40      &France         &Jul.\ 09, 1987 &Dec.\ 23, 2011&  6\,182\\
        DAX-30      &Germany        &Jan.\ 02, 1970 &Dec.\ 23, 2011& 10\,564\\
        SMI-20      &Switzerland    &Jan.\ 07, 1988 &Dec.\ 23, 2011&  5\,902
    \end{tabular}
    \caption{Description of the dataset used: time series of returns of daily closing prices of international stock indices.}
    \label{tab:index_data}
\end{table}
We illustrate some of the quantities introduced above on series of daily index returns. 
The properties of the time series used are summarized in Tab.~\ref{tab:index_data}.
\subsection{Conditional probabilities and 2-points dependences}
\begin{figure*}
    \center
    \subfigure[SP500]{\label{fig:SP500}\includegraphics[scale=.33,trim=0 0 0 25,clip]{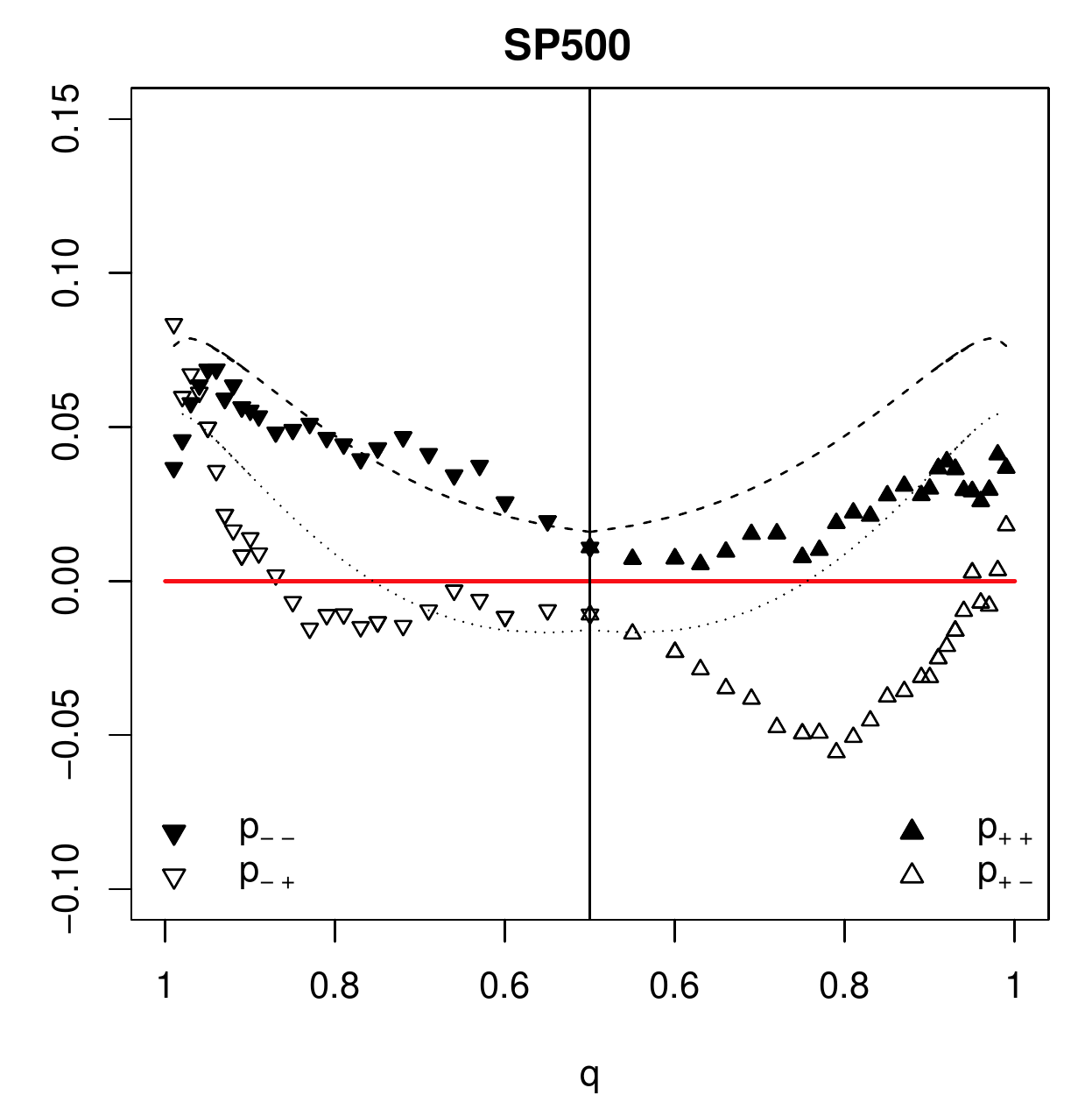}}%
    \subfigure[KOSPI]{\label{fig:KOSPI}\includegraphics[scale=.33,trim=0 0 0 25,clip]{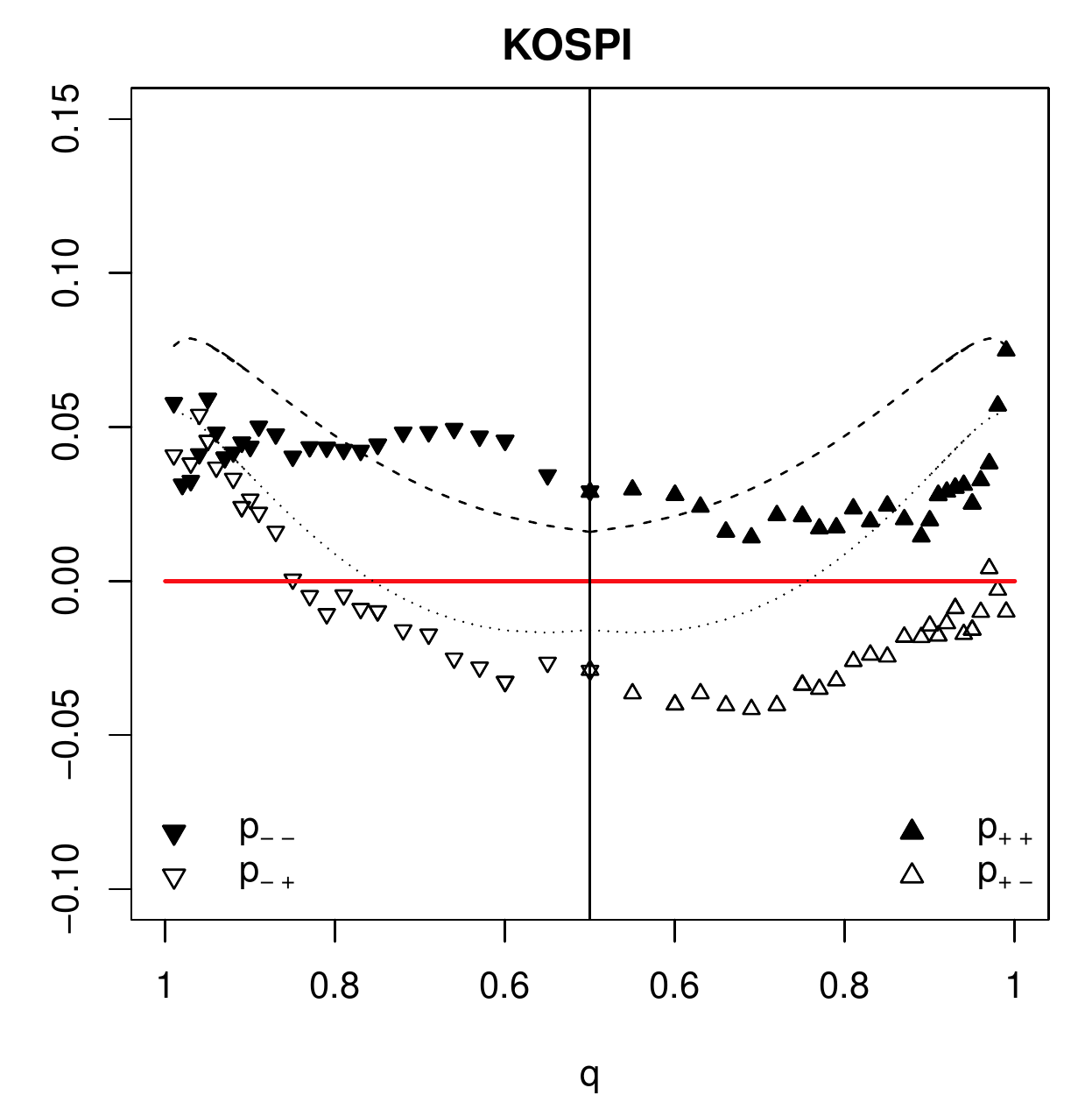}}
    \subfigure[CAC  ]{\label{fig:CAC}  \includegraphics[scale=.33,trim=0 0 0 25,clip]{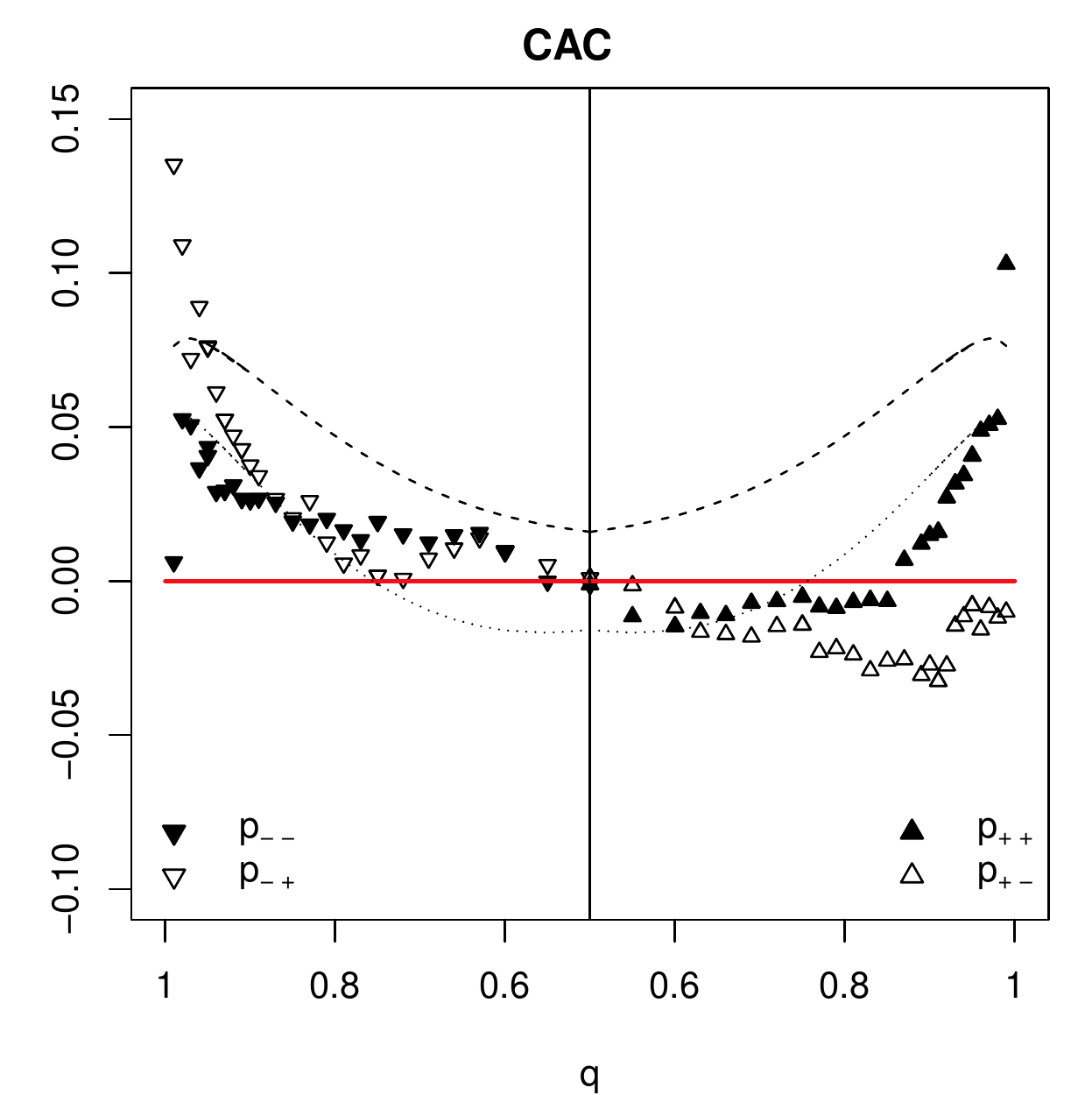}}\\
    \subfigure[DAX  ]{\label{fig:DAX}  \includegraphics[scale=.33,trim=0 0 0 25,clip]{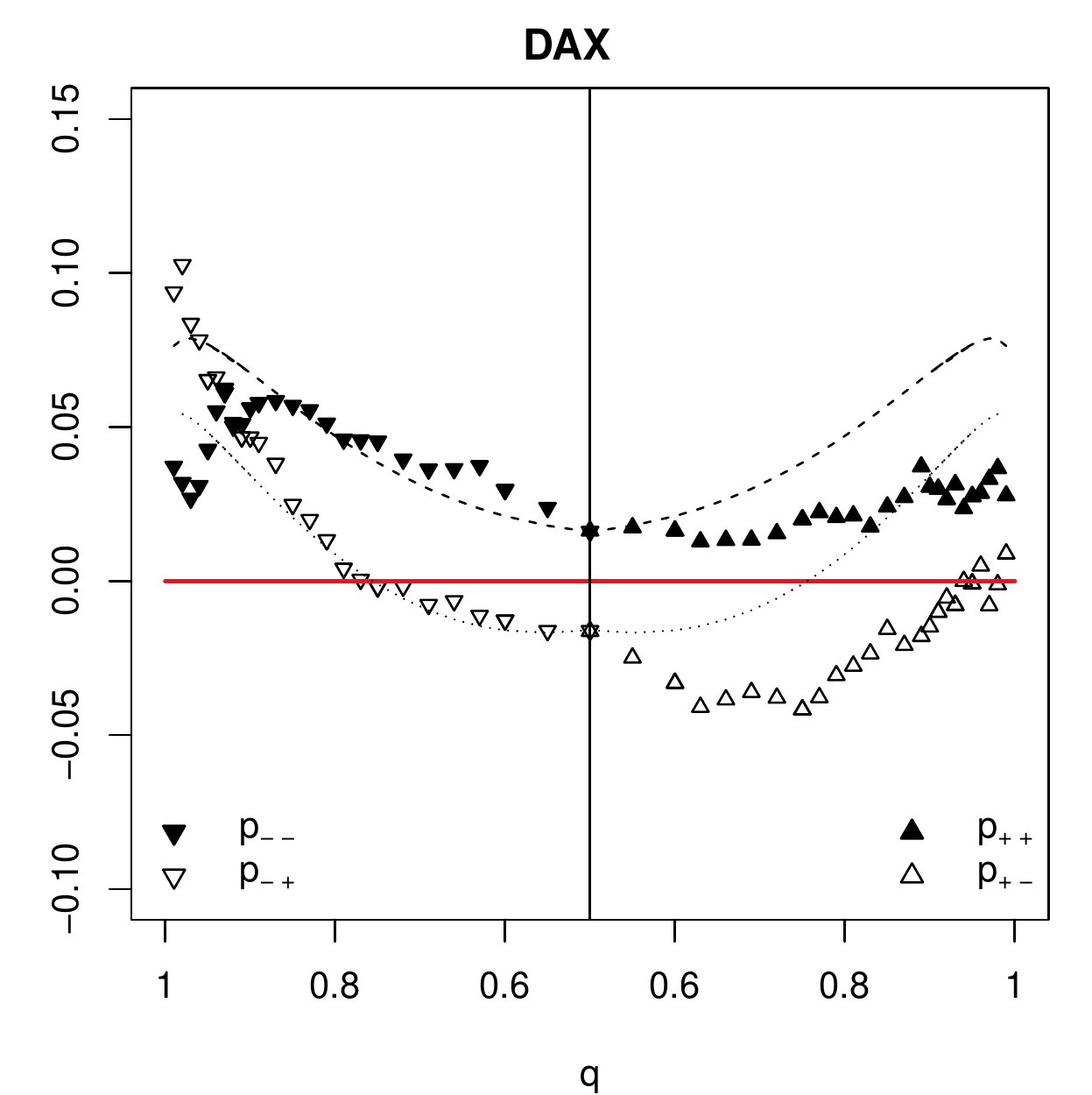}}
    \subfigure[SMI  ]{\label{fig:SMI}  \includegraphics[scale=.33,trim=0 0 0 25,clip]{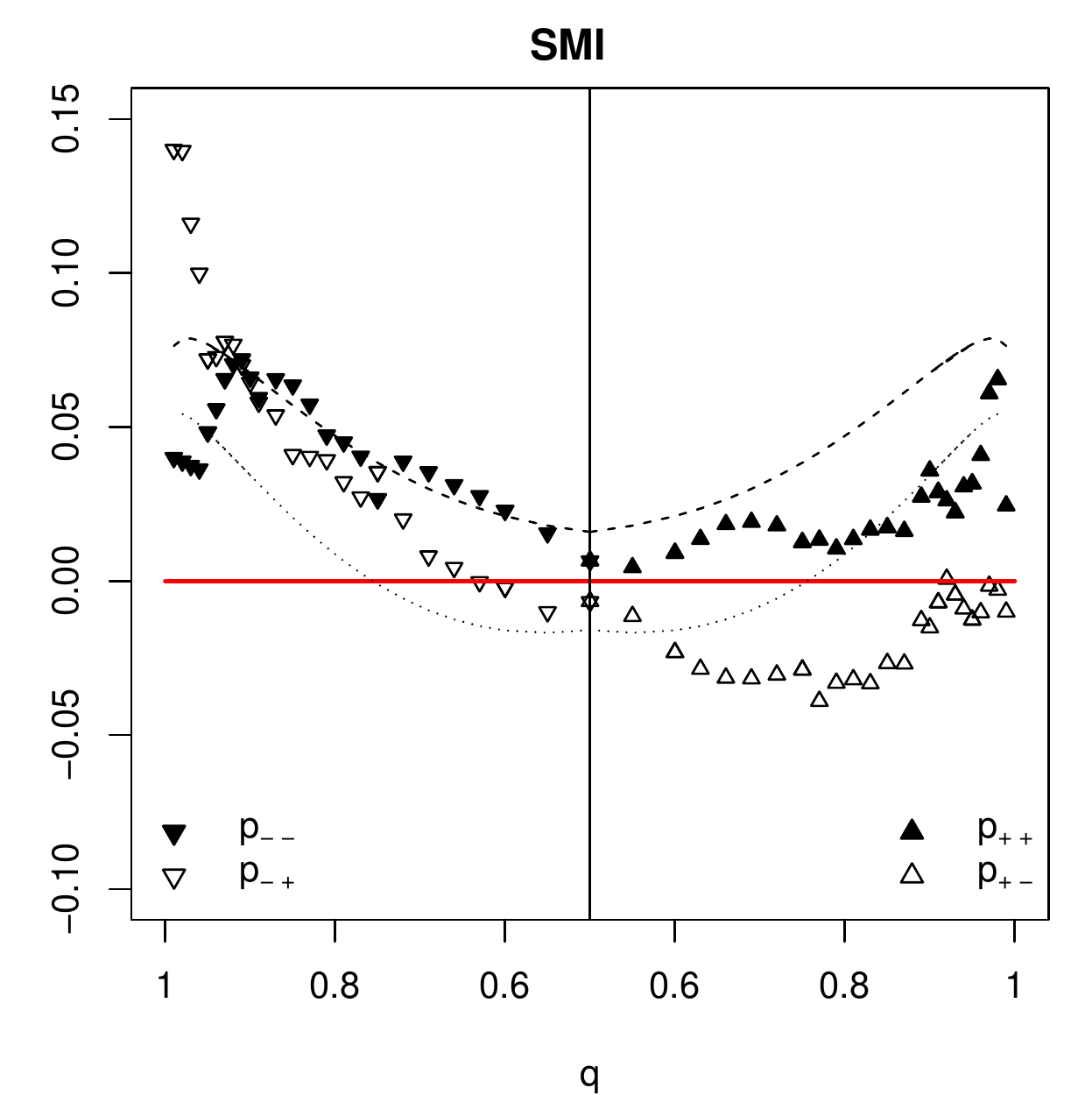}}
    \subfigure[EEG  ]{\label{fig:EEG}  \includegraphics[scale=.33,trim=0 0 0 25,clip]{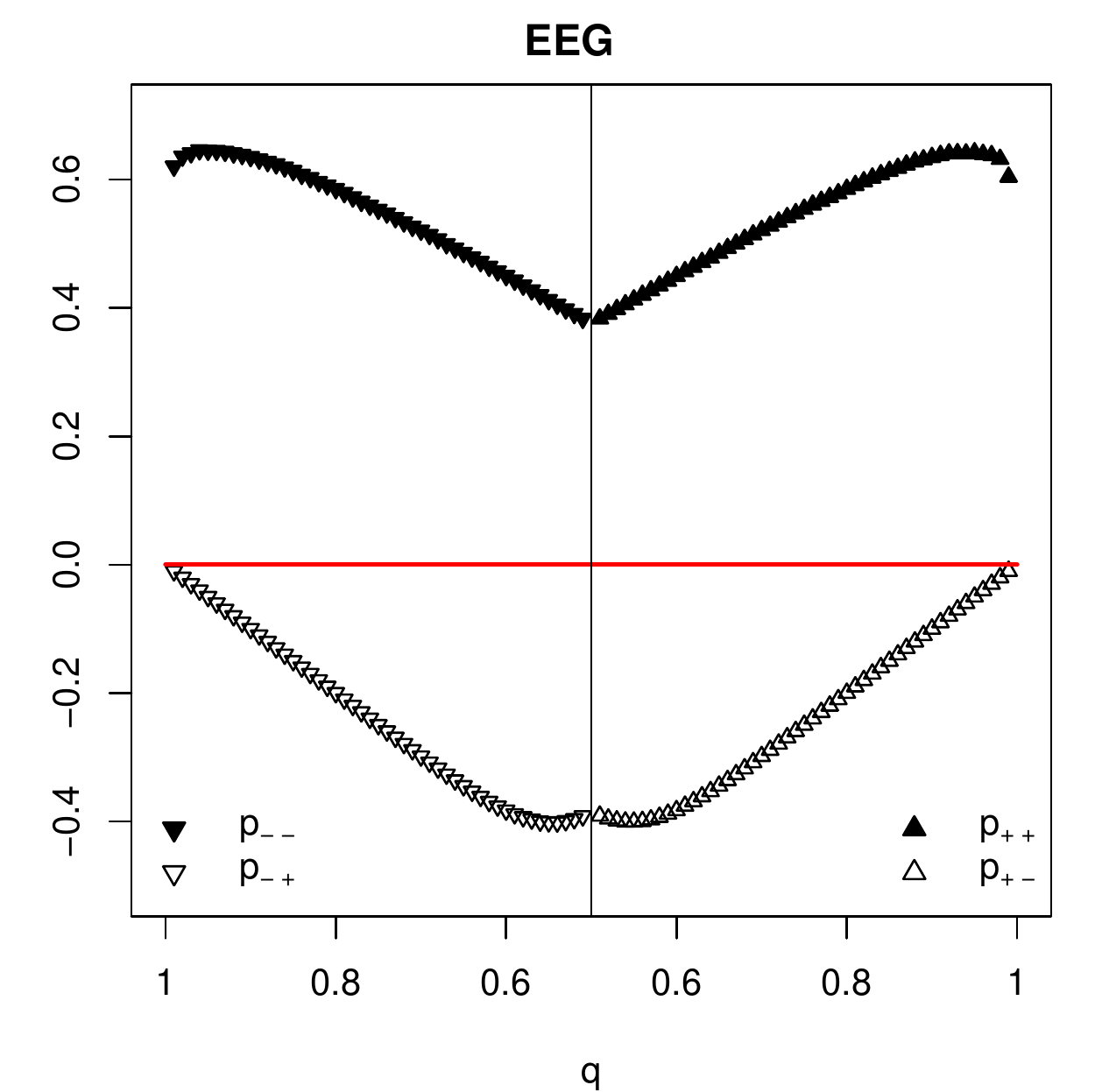}}
    \caption{Conditional extreme probabilities at $\ell=1$ (the WN contribution has been subtracted).
             Filled symbols are for persistence, and empty symbols for reversion.
             Upward pointing triangles are conditioned on positive jumps, and downward pointing triangles are conditioned on negative jumps.}
    \label{fig:condprob_idx}
\end{figure*}
We reproduce the study of \citet{boguna2004conditional} on the statistic of price changes conditionally on previous return sign,
and extend the analysis to any threshold $|\Xpm|\geq 0$ and to remote lags. 
In addition to the time series of the five stock indices presented in Tab.~\ref{tab:index_data}, we look at electroencephalogram (EEG) data from \cite{PhysRevE.64.061907}.
We first illustrate on Fig.~\ref{fig:condprob_idx} the conditional probabilities $p_{\pm\pm}^{(\ell)}$ (filled symbols)
and $p_{\pm\mp}^{(\ell)}$ (empty symbols) with varying threshold $q=F(\Xp)=1-F(\Xm)$, for $\ell=1$.
To study the departure from the independent case,
 it is more convenient to subtract the White Noise contribution, to get the corresponding \emph{excess} probabilities.

First, the EEG data, Fig.~\ref{fig:EEG}, exhibit a very strong and symmetric persistence;
reversion on the other side is shut down for extreme events (like for WN), and is more suppressed than WN for intermediate values. 
As of the plots relative to financial indices, several features can be immediately observed :
    positive events (upward triangles)   trigger more subsequent positive (filled) than negative (empty) events;
    negative events (downward triangles) trigger more subsequent negative (filled) than positive (empty) events, 
    except in the far tails $q\gtrsim 0.9$ where reversion is stronger than persistence after a negative event.
    Both these effects dominate the WN benchmark, but the latter effect is however much stronger.
This overall behavior is similar for the time series of returns of all the stock indices studied.
The shapes of $p_{\pm\pm}$ and $p_{\pm\mp}$ versus $q$ are not compatible with the Student copula benchmarks 
(correlation $\rho=0.05$ and d.o.f.\ $\nu=5$)
shown in dashed and dotted lines, respectively.
Notice that, due to its non-trivial tail-correlations, see Ref.~\cite{chicheportiche2012joint}, the Student copula 
does generate increased persistence with respect to WN, lower reversion in the core and higher reversion in the tails.
But empirically the reversion is asymmetric and typically stronger when conditioning on 
negative events rather than on positive events, a property reminiscent of the 
leverage effect which cannot be accounted for by a pure (symmetric) Student copula.
Some of the indices exhibit more pronounced reversion and persistence effects. 
Interestingly, the CAC-40 returns have a regime $0.5\leq q\lesssim 0.9$ close to a white noise 
(with, in particular, a value of $p_{\pm\pm}^{(1)}=p_{\pm\mp}^{(1)}$ very close to 0 at $q=0.5$,
revealing an inefficient conditioning, 
i.e.\ as many positive and negative returns immediately following positive or negative returns),
but the extreme positive events $q\gtrsim 0.9$ show a very strong persistence,
and the extreme negative events a very strong reversion.

\begin{figure}
\center
	\includegraphics[scale=.65,trim=0    0 1620 0,clip]{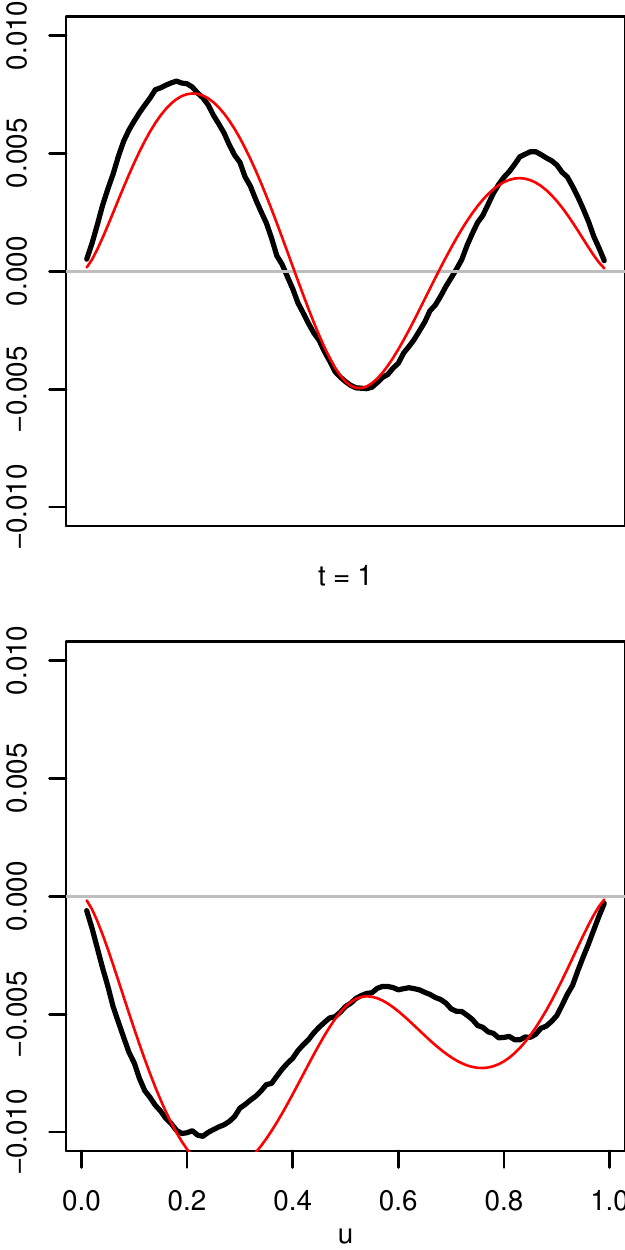}
	\includegraphics[scale=.65,trim=1275 0  350 0,clip]{images/fig4.pdf}
	\caption{Adaped from \cite{chicheportiche2011goodness}. 
    Diagonal (top) and anti-diagonal (bottom) of the self-copula for lags $\ell=1$ and $\ell=128$;
	the product copula has been subtracted. 
    The copula determined empirically on stock returns is in bold black, and 
    a fit with the model of \cite{chicheportiche2011goodness} is shown in thin red. }
    \label{fig:previewCop}
\end{figure}

\citet{chicheportiche2011goodness} study in detail the $p$- and $\ell$- dependence of 
$[\cop[\ell](p,p)-p^2]$ and $[\cop[\ell](p,1\!-\!p)-p\,(1\!-\!p)]$ 
--- which are straightforwardly related to $p_{\pm\pm}^{(\ell)}$ and $p_{\pm\mp}^{(\ell)}$, respectively --- 
and find that the self-copula
of stock returns can be modeled with a high accuracy by a log-normal volatility with log-decaying correlation, 
in agreement with multifractal volatility models. 
We give an overview of the results in Fig.~\ref{fig:previewCop}, 
for the aggregated copula of all stocks in the S\&P500 in 2000--2004.
It is possible to show precisely how every kind of dependence present in the underlying process (discussed in~\cite{perello2004multiple})
reflects itself in $p_{++}^{(\ell)}$ for different $q$'s:
short ranged linear anti-correlation accounts for the central part ($p\approx 0.5$) departing from the WN prediction,
long-ranged amplitude clustering is responsible for the ``M'' and ``W'' shapes that reveal excess persistence and suppressed reversion, 
and the leverage effect can be observed in the asymmetric heights of the ``M'' and ``W''.

\subsection{Recurrence intervals and many-points dependences}

\begin{figure}
    \center
    \includegraphics[scale=.6,trim=0 0 358 40,clip]{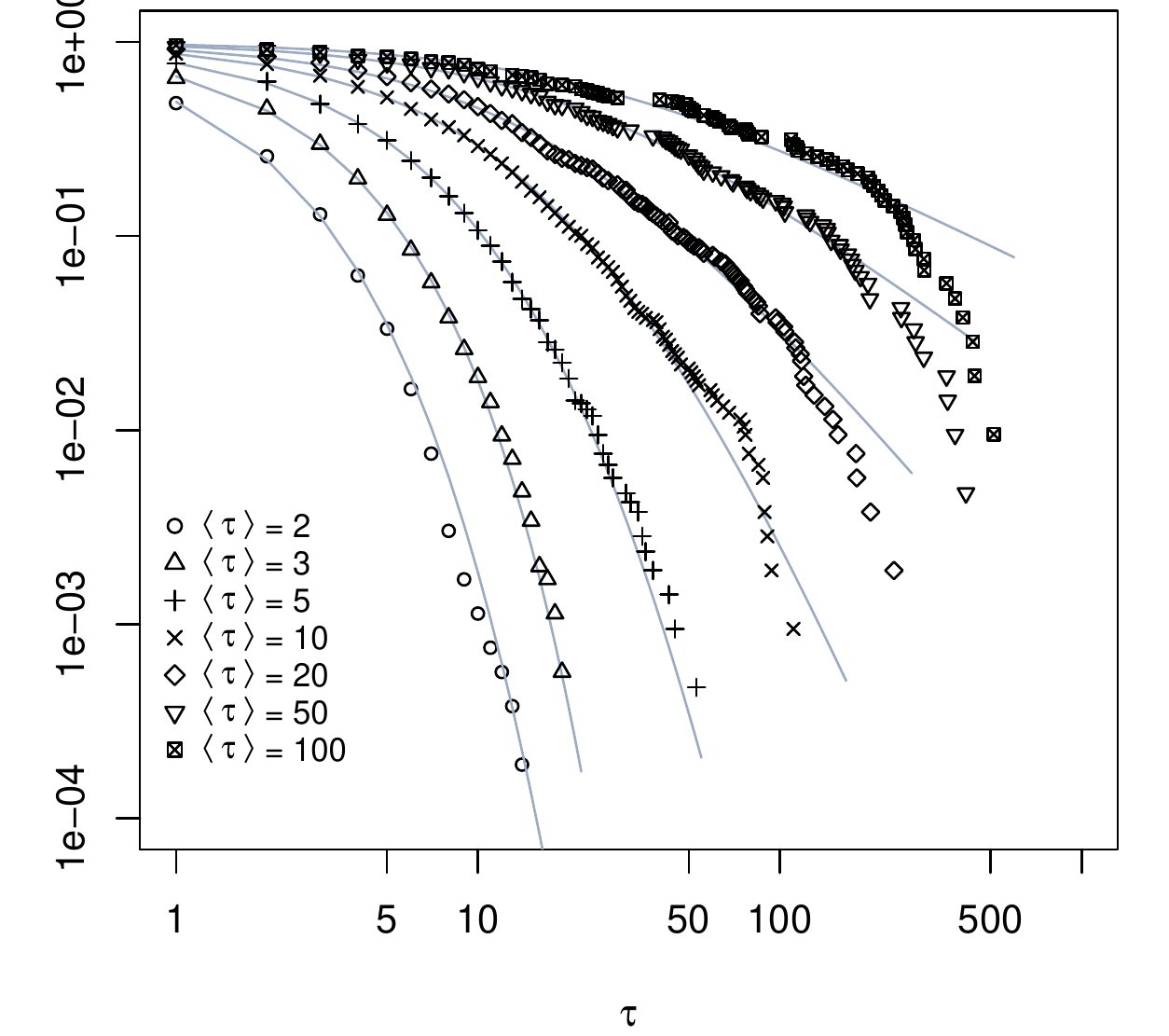}
    \includegraphics[scale=.6,trim=358 0 0 40,clip]{images/DAX_RI.pdf}
    \caption{DAX index returns. 
    \textbf{Top:} 
    tail probability $1-\Pi(\tau)$ of the recurrence intervals, at several thresholds $p_+=1/\vev{\tau}$,
    in log-log scale. 
    Grey curves are best fits to Eq.~\eqref{eq:ludescher-tsallis} suggested in Ref.~\cite{ludescher2011universal}.
    \textbf{Bottom:}
    estimated parameters $a$ and $b$ of the best fit.}
    \label{fig:RI_tCDF}
\end{figure}

Even the simple, two-points measures of self-dependence studied up to now show that
non-linearities and multi-scaling are two ingredients that must be taken into account
when attempting to describe financial time series;
we now examine their many-points properties.
As an example,
we compute the distribution of recurrence times of returns above a threshold $\Xp=F^{-1}(1\!-\!p_+)$.

Fig.~\ref{fig:RI_tCDF} shows the tail cumulative distribution 
 $1-\Pi(\tau)$ of the recurrence intervals of DAX returns, at several thresholds $p_+=1/\vev{\tau}$ 
 --- the distribution for other indices is very similar.
In the log-log representation used, an exponential distribution (corresponding to independent returns) 
would be concave and rapidly decreasing, while a power-law would decay linearly.
The empirical distributions fit neither of those, and \citet{ludescher2011universal}
suggested a parametric fit of the form 
\begin{equation}\label{eq:ludescher-tsallis}
    1-\Pi(\tau)=[1+b\,(a\!-\!1)\,\tau]^{(2\!-\!a)/(a\!-\!1)}.
\end{equation}
However, important deviations are present in the tail regions for thresholds at $\Xp\gtrsim F^{-1}(0.9)$, 
i.e.\ $\vev{\tau}\gtrsim 1/(1-0.9) =10$: as a consequence, there is no hope that 
the curves for different threshold collapse onto a single curve after a proper rescaling \cite{chicheportiche2013recurrences}, 
as is the case e.g.\ for seismic data.
A more fundamental determination of the form of $\Pi(\tau)$ should rely on Eq.~\eqref{eq:cumul_Pi}
and a characterization of  the $\tau$-points copula.

Similarly to the statistic of the recurrence intervals, their dynamics must be studied carefully. 
We have shown that the conditional distribution of recurrence intervals after a previous recurrence 
is very complex and involves long-ranged non-linear dependences, 
so that a simple assessment of recurrence times auto-correlation may not be informative enough for a
deep understanding of the mechanisms at stake.


\section{Discussion}

\subsection{Conditional Expected Shortfall}
\begin{figure}
\center
    \includegraphics[scale=.5]{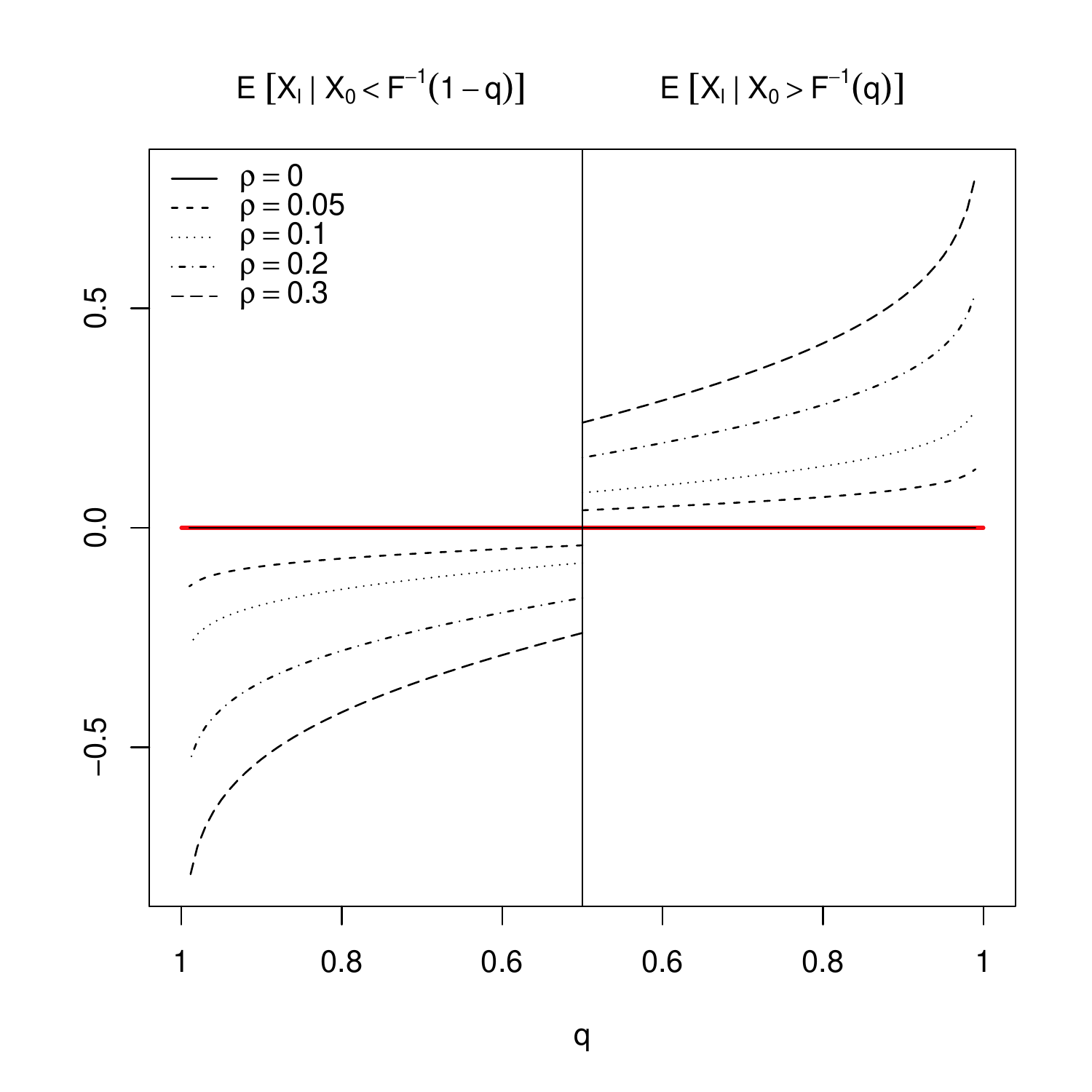}
    \caption{Conditional Expected Shortfall of a Gaussian pair $(X_0,X_\ell)$ for different values of $\rho(\ell)$.
             The value at $q=0.5$ is $\sqrt{{2}/{\pi}}\,\rho(\ell)$.}
    \label{fig:condESGauss}
\end{figure}

In addition to caring for frequencies of conditional events, 
one can try to characterize their magnitudes.
This of course does no longer fit in the framework of copulas (that ``count'' joint events) but
can instead be quantified by a multivariate generalization of the Expected Shortfall (or Tail Conditional Expectation).
For a single random variable with cdf $F$, the 
Expected Shortfall is the average loss when conditioning on large events:
\begin{align*}
    \text{ES}(p_-)&=\esp{X_t|X_t<\Xm }\\
                  &=\frac{1}{p_-}\int_{-\infty}^{F^{-1}(p_-)}x\,\dd{F(x)}\\
                  &=\frac{1}{p_-}\int_0^{p_-}F^{-1}(p)\,\dd{p}
\end{align*}
In the same spirit, for bivariate distributions, the mean return conditionally on preceding return `sign' is defined:
\begin{subequations}\label{eq:condES}
\begin{align}
    \vev{X}^{(\ell)}_-&= \esp{X_t|X_{t-\ell}<\Xm }\\
    \vev{X}^{(\ell)}_+&= \esp{X_t|X_{t-\ell}>\Xp }.
\end{align}
\end{subequations}
%
As an example, consider the Gaussian bivariate pair $(X_t,X_{t+\ell})$,
whose whole $\ell$-dependence is in the linear correlation coefficient $\rho(\ell)$.
Fig.~\ref{fig:condESGauss} shows the conditional Expected Shortfall that can be computed exactly from Eqs.~\eqref{eq:condES},
and is proportional to the inverse Mill ratio:
\[
    \vev{X}_\pm=\pm\rho(\ell)\frac{\Phi'(\Xpm)}{p_\pm},
\]
where $\Phi$ denotes the CDF of the univariate standard normal distribution.

This Gaussian prediction is to be compared with an empirical assessment of the same quantity
for series of returns of stock indices. 
Fig.~\ref{fig:condES} displays the behavior of $\vev{X}_\pm$ versus $q$ (we also show the median $\operatorname{med}(X)_\pm$) at 
lags corresponding to one day ($\ell=1$), one week ($\ell=5$) and one month ($\ell=20$).
The conditional amplitudes $\vev{X}_\pm$ measure ``how large'' a realization will be on average
after an event at a given threshold, 
whereas the conditional probabilities $p_{\pm\pm}$ and $p_{\pm\mp}$ quantify ``how often''
repeated such events occur.
Mind the \emph{unconditional} mean and median values, both above zero and distinct from each other.
At $\ell=1$, the reversion of \emph{extreme} events is revealed again by the change of monotonicity from $q\approx 0.8$ on, 
and more strongly for $q>0.9$ where $\vev{X}_-$ has an opposite sign than the preceding return;
this corroborates the observation made on conditional probabilities above.
Beyond the next day, the general picture is that dependences tend to vanish 
and the empirical measurements get more concentrated around the WN prediction.
However, tail effects are strongly present, with unexpectedly a typical behavior opposite
to that of $\ell=1$: weekly, monthly reversion of extreme positive jumps. 
See the caption for a detailed discussion of the specificities of each stock index at every lag $\ell$.
\begin{figure*}
    \center
    \subfigure[SP500]{\label{fig2:SP500}\includegraphics[scale=.575]{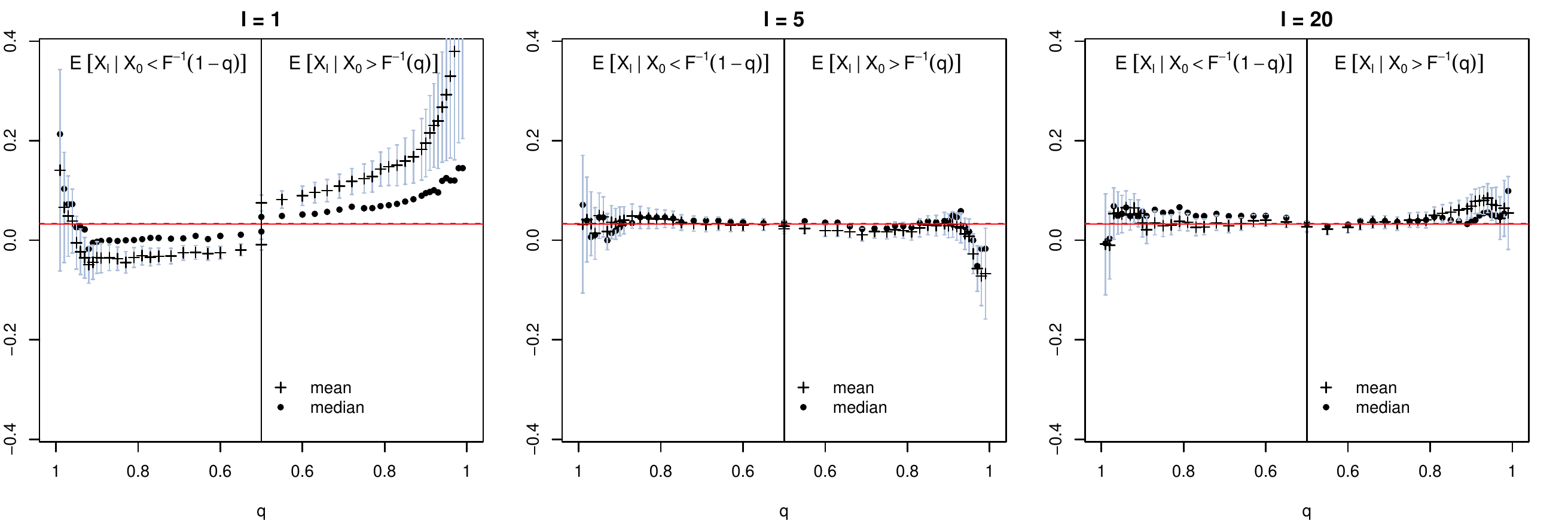}}
    \subfigure[KOSPI]{\label{fig2:KOSPI}\includegraphics[scale=.575]{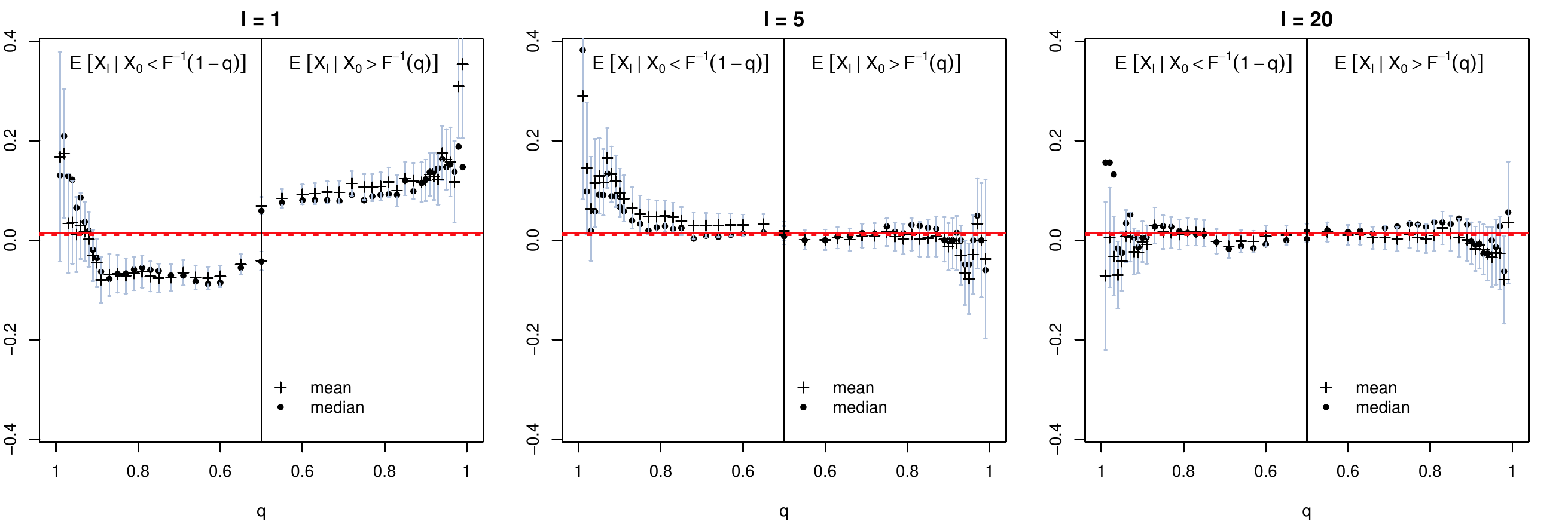}}
    \caption{Conditional extreme amplitudes, at lags $\ell=1, 5, 20$.
             The upper-right and lower-left quadrants express persistence,
             while the upper-left and lower-right quadrants reveal reversion.
             For a scale-free dependence structure, one would expect the magnitudes to decrease with the lag $\ell$
             but the global shape to be conserved.
             What we instead observe is important changes of configuration at different lags:
             For example, the strong reversion of negative tail events at $\ell=1$ vanishes
             at farther lags, and even turns into strong persistence for the CAC and DAX indices.
             That is to say that these indices tend to mean-revert after a negative event at the daily frequency,
             but to trend on the weekly scale.
             Similarly, the strong persistence of positive events at $\ell=1$ converts to a strong reversion 
             in the tails at $\ell=20$ for the European indices (CAC, DAX); 
             a weaker reversion is observed at intermediate scale ($\ell=5$) for most indices (including US and Korean).
             }
    \label{fig:condES}
\end{figure*}
\begin{figure*}
    \addtocounter{figure}{-1}
    \addtocounter{subfigure}{2}
    \center
    \subfigure[CAC  ]{\label{fig2:CAC}  \includegraphics[scale=.5]{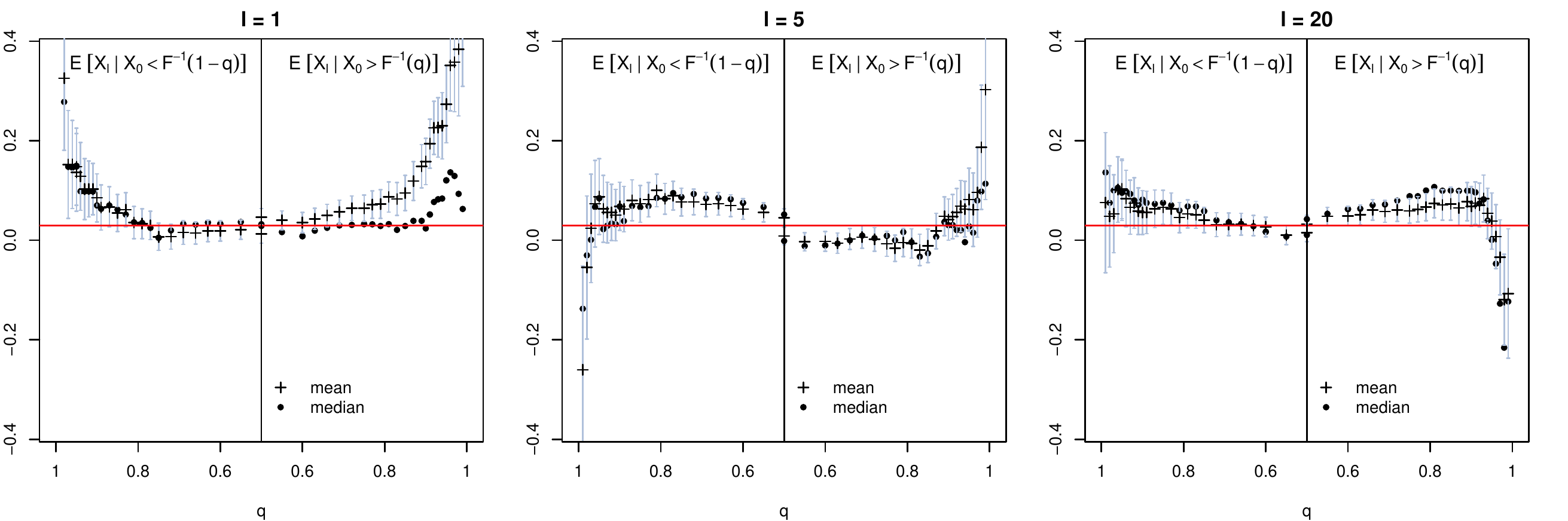}}
    \subfigure[DAX  ]{\label{fig2:DAX}  \includegraphics[scale=.5]{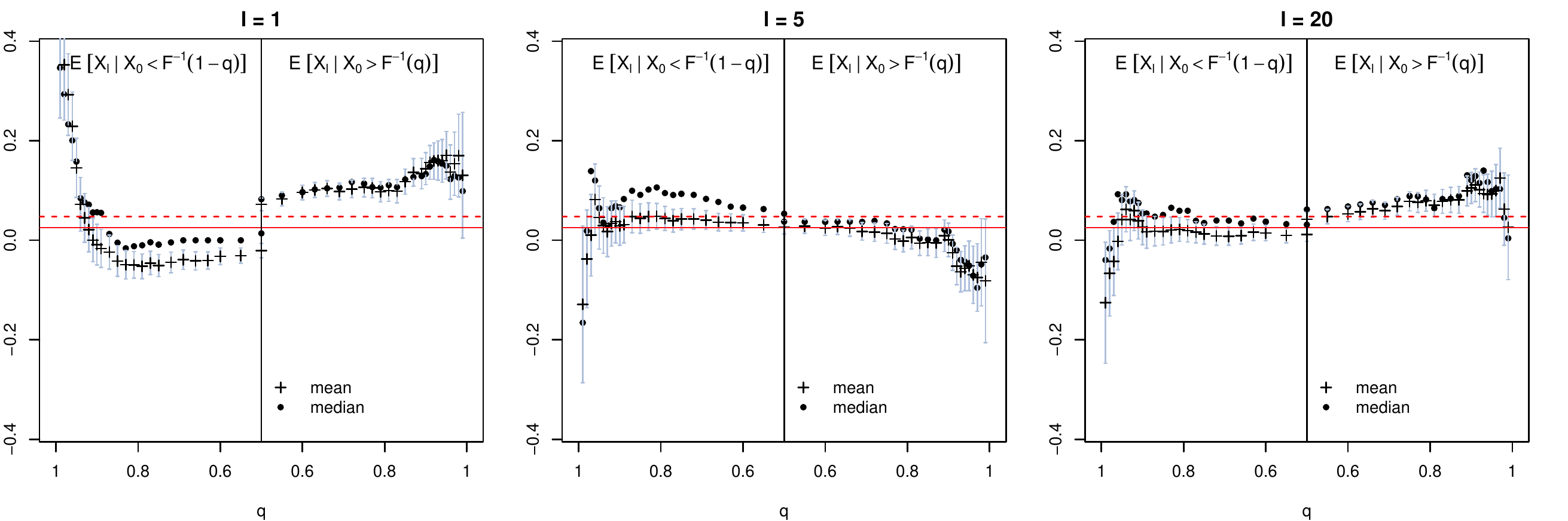}}
    \caption{(continued)}
\end{figure*}

\subsection{Conclusion}
We report several properties of recurrence times and the statistic of other 
observables (waiting times, cluster sizes, records, aftershocks) in light of their description in terms of the diagonal copula,
and hope that these studies can shed light on the $n$-points properties of the process by 
assessing the statistics of simple variables
rather than positing an {\it a priori} dependence structure.
 
The exact universality of the mean recurrence interval imposes a natural scale in the system.
A scaling relation in the distribution of such recurrences is only possible in absence of any other characteristic time.
When such additional characteristic times are present (typically in the non-linear correlations), 
no such scaling is expected, in contrast with time series of earthquake magnitudes.

We also stress that recurrences are intrinsically multi-points objects 
related to the non-linear dependences in the underlying time-series.
As such, their autocorrelation is not a reliable measure of their dynamics,
for their conditional occurrence probability is much history dependent.

Ultimately, recurrences may be used to characterize risk in a new fashion.
Instead of --- or in addition to --- caring for the amplitude and probability of adverse events at a given horizon,
one should be able to characterize the risk in a dynamical point of view.
In this sense, an asset $A_1$ could be said to be ``more risky'' than another asset $A_2$
if its distribution of recurrence of adverse events has such and such ``bad'' properties that $A_1$ does not share.
This amounts to characterizing the disutility  by ``When?''  shocks are expected to happen,
in addition to the usual ``How often?'' and ``How large?''.


It would be interesting to study many-points dependences in  continuous-time proceses,
where the role of the $n$-points copula is played by a counting process.
The events to be counted can either be triggered by an underlying continuous process crossing a threshold, 
or more directly be modeled as a self-exciting point process, like a Hawkes process.
A typical financial application could be found in transaction times in a Limit Order Book.

\acknowledgments
We thank F.~Abergel, D. Challet and G.~Tilak, for helpful discussions and comments. 
R.~C. acknowledges financial support by Capital Fund Management, Paris.

\appendix
\section{Simple copulas and Sklar's theorem}
Sklar's theorem \cite{sklar1959fonctions} states that 
any multivariate distribution $\mathcal{F}_{\libracket 1,n\ribracket}(x_1,\ldots,x_n)$ can be written in terms of 
univariate marginal distribution functions $F_i(x_i)$  ($i=1,\ldots,n$) and a `copula' function $\cop(u_1,\ldots,u_n)$ on $[0,1]^n$ which, 
by definition, characterizes the dependence structure between the variables.
In practice, constructing the copula is achieved letting $u_i=F_i(x_i)$ for every variable $i$.
This is expressed mathematically by Eq.~\eqref{eq:biv_cop} for bivariate distributions, 
and can be generalized straightforwardly (see Eq.~\eqref{eq:def_cop_n} for the \emph{diagonal} of the $n$-points copula).

As an example, the Gaussian diagonal copula is
\begin{equation}\label{eq:GaussCop}
    \CopN[n]{p}=\Phi_\rho\big(\Phi^{-1}(p),\ldots,\Phi^{-1}(p)\big)
\end{equation}
where $\Phi^{-1}$ is the univariate inverse CDF, and $\Phi_\rho$
denotes the multivariate CDF with $(n\times n)$ covariance matrix $\rho$,
which is T\oe{}plitz with symmetric entries
\begin{equation}
    \rho_{tt'}=\rho(|t-t'|),\quad t,t'=1,\ldots,n.
\end{equation}
The White Noise (WN) product copula
 $\CopN[n]{p}=p^n$ is recovered in the limit of vanishing correlations $\rho(\ell)=\delta_{\ell 0}$, and
other examples include the exponentially correlated Markovian Gaussian noise, 
the logarithmically correlated multi-fractal Gaussian noise,
and the power-law correlated (thus scale-free) fractional Gaussian noise.

\begin{figure}
    \center
    \includegraphics[scale=.55,trim=0 0 350 0,clip]{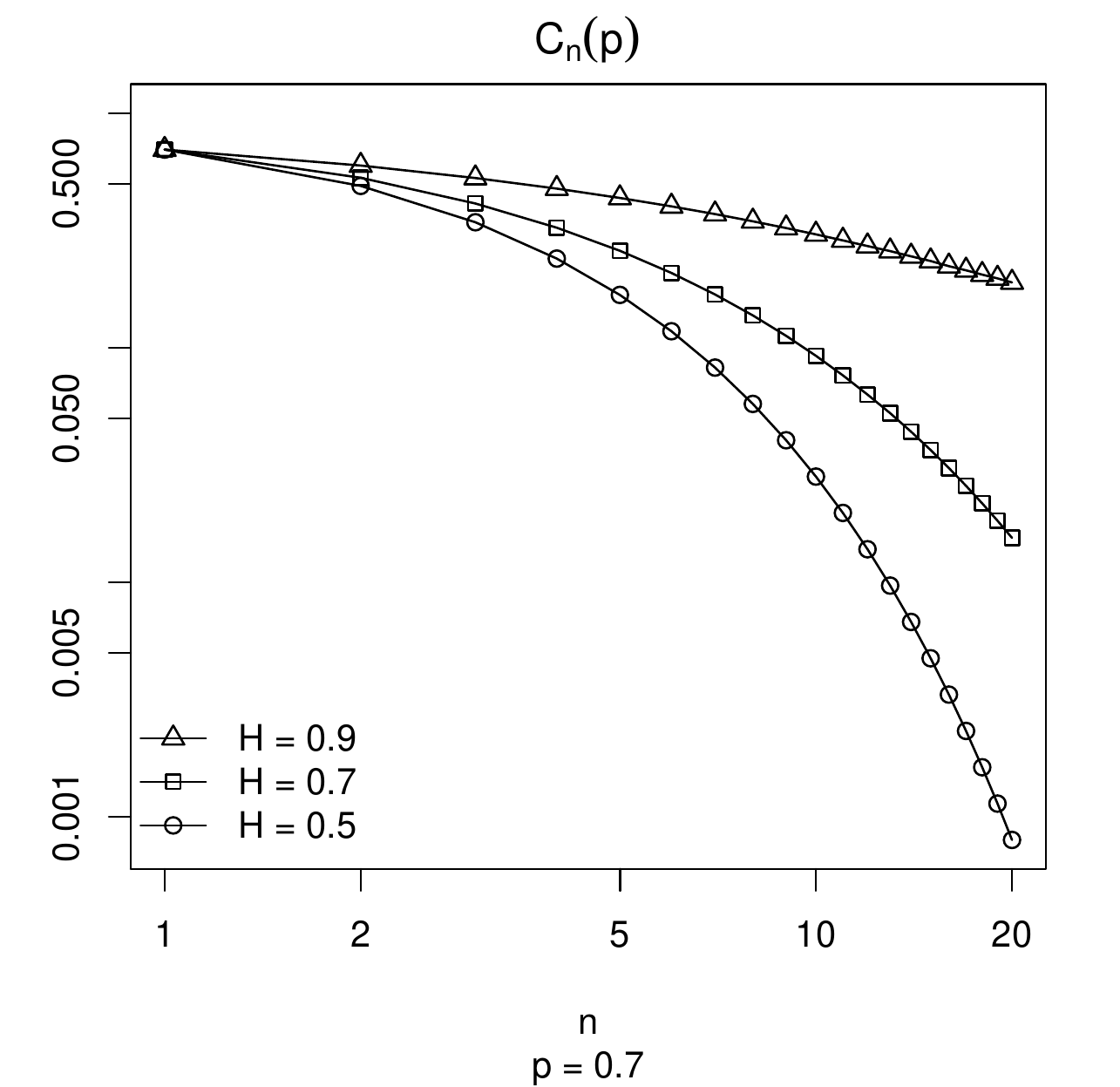}
    \caption{Copula of the Fractional Gaussian Noise (FGN) with different Hurst exponents, and at threshold $p_+=0.7$ (log-log scale).}
    \label{fig:gausscop}
\end{figure}

 Fig.~\ref{fig:gausscop} displays $C_n(p_+\!=\!0.7)$ versus $n$ for different Hurst indices $H=0.5,0.7,0.9$.
 The asymptotic behaviour at large  $n$ cannot be displayed here because of numerical restrictions, 
 but the small $n$ properties are more relevant for characterizing short-time conditional dynamics.
 
\bibliography{Chicheportiche-Chakraborti-long}

\end{document}